\documentclass{aa}

\usepackage{graphicx}
\usepackage{txfonts}
\usepackage{units}
\usepackage{url}
\usepackage{natbib}
\usepackage{units}

\newcommand{\lsi}{LS~I~$+61^{\circ}303$}

\newcommand{\beq}{\begin{equation}}
\newcommand{\eneq}{\end{equation}}

\newcommand{\fermilat}{\textit{Fermi}-LAT}

\begin{document} 

\title{Understanding the periodicities in radio and GeV emission
  from \lsi{}}
\author{
  F.~Jaron\inst{1}
  \and 
  G.~Torricelli-Ciamponi\inst{2}
  \and
  M.~Massi\inst{1}
}

\institute{
  Max-Planck-Institut f\"ur Radioastronomie, 
  Auf dem H\"ugel 69, 
  53121 Bonn, Germany\\
  e-mail: \url{fjaron, mmassi@mpifr-bonn.mpg.de}
  \and
  INAF - Osservatorio Astrofisico di Arcetri, L.go E. Fermi 5,
  Firenze, Italy\\ 
  e-mail: \url{torricel@arcetri.astro.it}
}

\date{Received ???; accepted ???}

\abstract{ 
  One possible scenario to explain the emission from the stellar
  binary system \lsi{} is that the observed flux is emitted by
  precessing jets powered by accretion. Accretion models predict two
  ejections along the eccentric orbit of \lsi{}: one major ejection at
  periastron and a second, lower ejection towards apastron. Our GeV
  gamma-ray observations show two peaks along the orbit (orbital
  period~$P_1$) but reveal that at apastron the emission is also
  affected by a second periodicity, $P_2$. Strong radio outbursts also
  occur at apastron, which are affected by both periodicities
  (i.e.\ $P_1$ and $P_2$), and radio observations show that $P_2$ is
  the precession of the radio jet. Consistently, a long-term
  modulation, equal to the beating of $P_1$ and $P_2$, affects both
  radio and gamma-ray emission at apastron but it does not affect
  gamma-ray emission at periastron.
}{ 
  If there are two ejections, why does the one at periastron not
  produce a radio outburst there? Is the lack of a periastron radio
  outburst somehow related to the lack of $P_2$ from the periastron
  gamma-ray emission?
}{ 
  We develop a physical model in which relativistic electrons are
  ejected twice along the orbit. The ejecta form a conical jet that is
  precessing with $P_2$. The jet radiates in the radio band  by the
  synchrotron process and the jet radiates in the GeV energy band  by the
  external inverse Compton and synchrotron self-Compton  processes.
  We compare the output fluxes of our physical model with two
  available large archives:  Owens Valley Radio Observatory
    (OVRO) radio and  \textit{Fermi} Large Area Telescope (LAT)
  GeV observations,  the two databases overlapping for five
    years.
}{ 
  The larger ejection around periastron passage results in a slower
  jet, and severe inverse Compton losses result in the jet also being
  short. While large gamma-ray emission is produced, there is only
  negligible radio emission. Our results are that the periastron jet
  has a length of $3.0 \cdot 10^6 {\rm r_s}$ and a velocity $\beta
  \sim 0.006,$ whereas the jet at apastron has a length of $6.3 \cdot
  10^7 {\rm r_s}$ and $\beta \sim 0.5$.
}
{ 
  In the accretion scenario the observed periodicities can be
  explained if the observed flux is the intrinsic flux, which is a
  function of $P_1$, times the Doppler factor, a function of
  $\beta\cos(f(P_2))$. At periastron, the Doppler factor is scarcely
  influenced by $P_2$ because of the low $\beta$. At apastron the
  larger $\beta$ gives rise to a significant Doppler factor with
  noticeable variations induced by jet precession.
}

\keywords{
  Radio continuum: stars - X-rays: binaries - X-rays: individual
  (\lsi{}) - Gamma-rays: stars
}

\maketitle

\section{Introduction}
\label{sect:introduction}

\begin{figure*}
  \hspace{.1\textwidth}
  \includegraphics[]{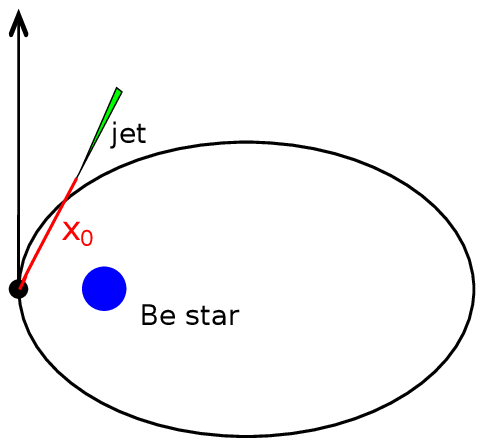}
  \hspace{.2\textwidth}
  \includegraphics[]{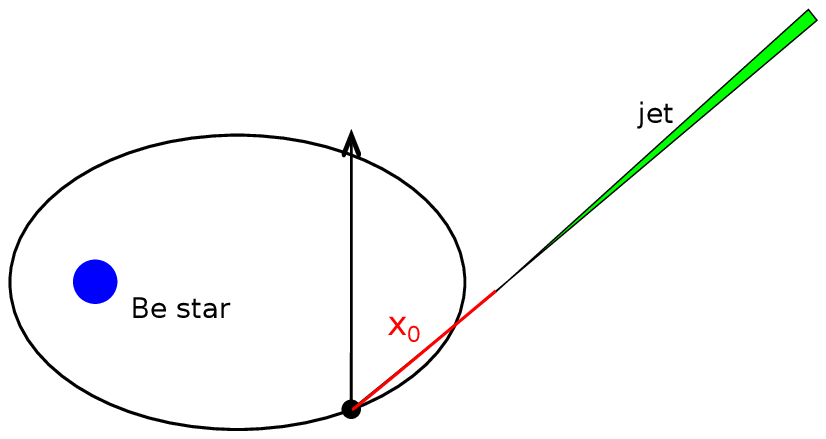}
  \hspace{.03\textwidth}\\
  \includegraphics{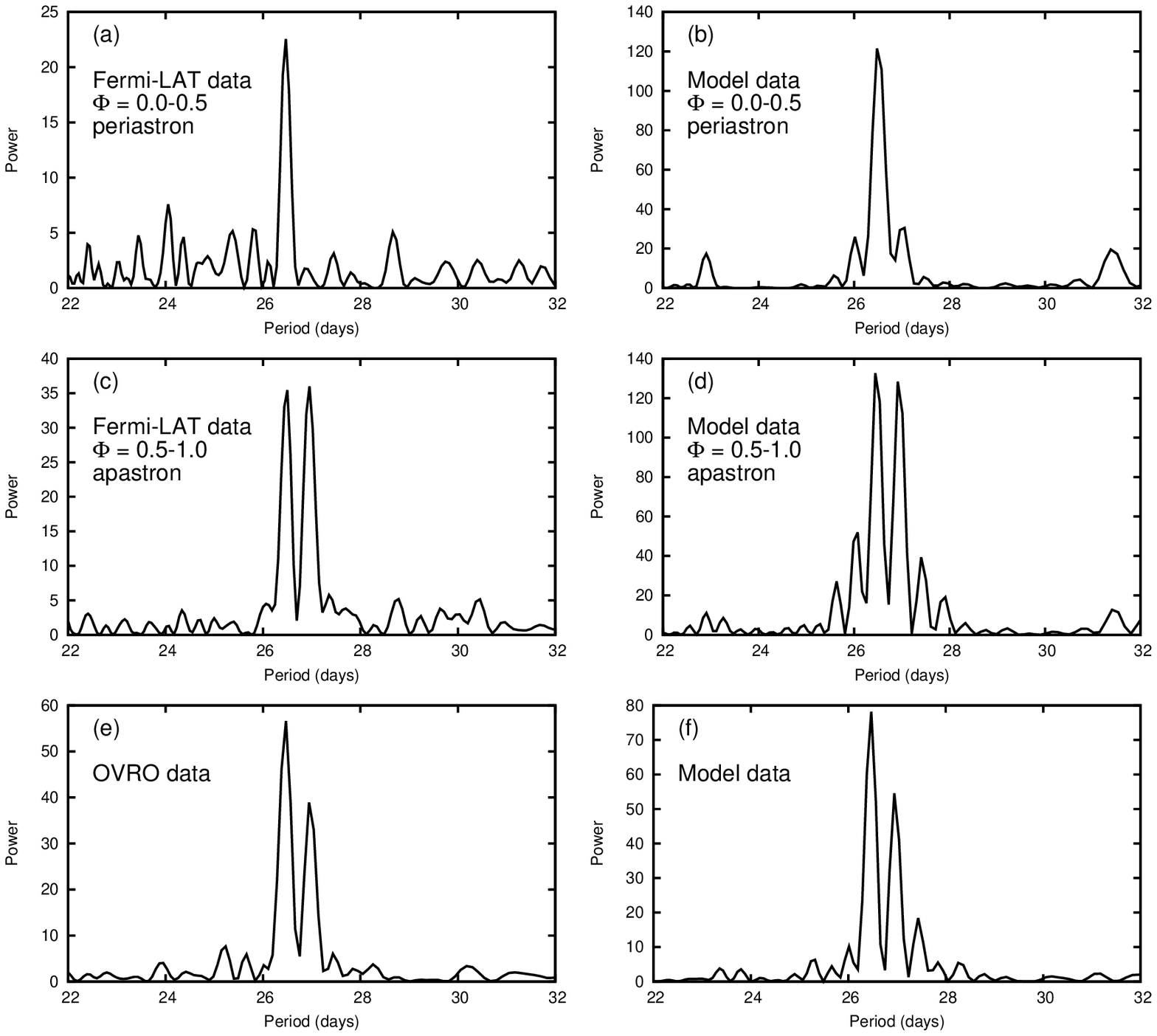}
  \caption{
    Top:  Sketch of the stellar system\ \lsi{}. We assume two
    ejections of relativistic electrons along the orbit as in
    \citet{Bosch-Ramon2006}. The two ejections refill the conical
    precessing jet defined in \citet{Massi2014}. The electrons emit by
    synchrotron and  IC processes. The resulting jet parameters
    are given in  Table 2. The ejection at periastron gives rise
    to gamma-ray emission only and the periodogram of this emission
    only shows $P_1$, the orbital period. The ejection towards
    apastron is associated with gamma-ray and radio emission. The
    periodograms for both emissions shows two features, at $P_1$ and
    $P_2$, the precession period of the jet.
    (a) Lomb-Scargle periodogram for \fermilat{} data from the orbital
    phase interval\ $\Phi = 0.0-0.5$ (periastron). The orbital
    period\ $P_1$ is clearly detected.
    (b) Lomb-Scargle periodogram of the model gamma-ray data (IC) from
    the orbital phase interval\ $\Phi = 0.0-0.5$ (periastron). The
    orbital period\ $P_1$ is the only peak, in agreement with the
    observations.
    (c) In the orbital phase interval\ $\Phi = 0.5-1.0$ (apastron),
    there is a peak at the precession period\ $P_2$ in addition to the
    orbital period\ $P_1$.
    (d) The two peaks at $P_1$ and $P_2$ in the apastron data are
    reproduced by the model.
    (e) Lomb-Scargle periodogram for OVRO radio data for the whole
    orbital phase interval.  Both $P_1$ and $P_2$ are present.
    (f) Lomb-Scargle periodogram of the model radio data (synchrotron
    emission). Both peaks are reproduced by the model.
  }
  \label{fig:timing}
\end{figure*}

The emission from the X-ray binary \lsi{} has been observed from radio
wavelengths to $\gamma$-rays \citep{Taylor1982, Paredes1994,
  Mendelson1989, Zamanov1999, Harrison2000, Abdo2009}. Different
variability patterns present in the observed flux at different
energies constitute both a tool and a challenge for every possible
interpretation of the physical processes behind the emission from this
source. In this binary system, the compact object (neutron star or
black hole) travels around a fast rotating Be star (see sketch in
Fig.~\ref{fig:timing}). \citet{Casares2005}, who used only He~I and
He~II lines in the spectral range $3850 - 5020$~\AA{} to avoid
contamination from the emission lines of the decretion disk of the Be
star, determined the phase of periastron as $\Phi = 0.23 \pm 0.02$ and
an eccentricity of $e = 0.72 \pm 0.15$.

The specific nature of the interaction between the matter expelled
from the Be star and the compact object defines the origin and physics
of the observed emission; different models have been proposed assuming
that either the compact object is accreting \citep{Taylor1992,
  Marti1995, Bosch-Ramon2006, Romero2007, Massi2014} or non-accreting
\citep{Maraschi1981, Chernyakova2006, Dubus2006}. The challenge of all
models is their capability of reproducing both the observed emission
over the entire spectrum and the temporal variability specific of each
energy range. In fact, the observed periodicities in different energy
bands constitute a useful test for the validity of each model and
provide the opportunity to allow the determination of the emission
origin in \lsi{}.

\citet{Massi2014} have developed a model that fits the observed radio
emission from \lsi{} and its periodicities. In this model the radio
emission is due to a precessing ($P_2$) jet that is periodically
($P_1$) refilled with relativistic particles. In this paper we further
test our model by extending its capabilities to reproduce the GeV
gamma-ray emission as well. The emission, for both radio and gamma-ray
energy bands, is produced by energetic particles ejected twice along
the orbit of \lsi,{} as in the context of accretion models.

The \citet{Bondi1944} accretion rate $\dot{M} \propto
\frac{\rho}{v^3}$ onto a compact object,  in addition to a main
peak around periastron (where the density\ $\rho$ has its maximum),
already starts to develop a second accretion peak towards apastron for
an eccentricity of $e = 0.4$ \citep{Taylor1992}. The reason for the
second accretion peak towards apastron is the lower relative
velocity\ $v$ between the accretor and the Be star wind, which
compensates for the lower density of the wind \citep{Taylor1992,
  Marti1995, Bosch-Ramon2006, Romero2007}. Indeed, gamma-ray emission
in the GeV regime, observed by the \textit{Fermi} Large Area Telescope
(LAT), occurs both towards periastron and apastron. Radio outbursts,
however, occur only towards apastron. These radio outbursts are
affected by a long-term modulation (timescale years)
\citep{Gregory2002} and timing analysis results in two strong spectral
features, $P_1$ and $P_2$, both of the order of one month
\citep{Massi2013, Massi2015, Massi2016}. Concerning the GeV gamma-ray
emission, timing analysis reveals that whereas the apastron emission
is modulated by $P_1$ and $P_2$,  identical to the radio 
  emission, the periastron GeV emission is only modulated by $P_1$
\citep{Jaron2014}. \citet{Ackermann2013} find that the GeV emission at
apastron is affected by  the long-term modulation ( as the
radio emission), while the periastron emission is not.

The aim of this paper is to investigate the physical processes
responsible for the radio and gamma-ray emission and to explain the
absence of $P_2$ in the emission at high energy around periastron as
well as the lack of a periastron radio outburst. In
Sect.~\ref{sect:scenarios}, we report on the observational results
revealing the nature of the physical processes behind the two
periodicities $P_1$ and $P_2$. The model is described in
Sect.~\ref{sect:methods} as a conical jet, precessing with period $P_2$,
periodically ($P_1$) refilled with relativistic electrons twice along
the orbit, and embedded in the photon fields of the companion
star. The emission of the electrons is due to synchrotron and inverse
Compton  (IC) processes. We compare the model with a more than
five-year overlapping monitoring of gamma-ray and radio emission, by
\fermilat{} and the Owens Valley Radio Observatory (OVRO), whose
calibration and reduction are presented in
Sect.~\ref{sect:observations}. In Sect.~\ref{sect:results} we present
our results and in Sect.~\ref{sect:conclusions} we give our
conclusions.

\section{The scenarios for \lsi{} and its periodicities}
\label{sect:scenarios}

In this section we report on observational evidence for a precessing
($P_2$) jet (Sects.~\ref{sect:precessingjet} and \ref{sect:beating})
and for two periodic ($P_1$) ejections along the orbit of \lsi{}
(Sect.~\ref{sect:Bedisk}).

\subsection{The precessing ($P_2$) jet}
\label{sect:precessingjet}

Interferometric images of \lsi{} at high resolution have at some
epochs shown a one-sided jet structure and at other epochs a two-sided
structure \citep{Massi1993, Massi2002, Taylor2002,
  Massi2004}. \citet{Dubus2006} developed a numerical model for a
pulsar nebula in which shocked material, due to the interaction of the
relativistic pulsar wind with the stellar wind from the companion
\citep{Maraschi1981, Chernyakova2006}, flows away in a comet-shape
tail, i.e. compatible  with a one-sided structure. \citet{Dhawan2006}
performed a set of VLBA observations and related the observed varying
morphology of the images to the rapid changes of a cometary tail. The
same VLBA data, re-processed in \citet{Massi2012}, confirm rapid
changes in position angle and that the radio structure is at some
epochs two-sided and at other epochs one-sided as expected from a jet
of a microquasar with variable ejection angle \citep{Massi2004}. The
compact jet associated with microquasars has a flat  or even
inverted radio spectrum \citep{Fender1999}. A flat  or inverted
radio spectrum for \lsi{} was already proven in the past
\citep{Gregory1979, Massi2009} and was recently measured over the
entire cm radio band (up to 9~mm) by \citet{Zimmermann2015}.

As stated above, in the context of microquasars the changing position
angle of the jet for \lsi{} is attributed to a variation of the angle
$\eta$ between the jet and line of sight \citep{Massi2004,
  Massi2009}. Such a change gives rise to a continuously changing
Doppler factor (the Doppler factor depending on the angle $\eta$) and
therefore to a changing morphology. For small angles the counter-jet
gets strongly attenuated and only the boosted approaching jet appears,
giving rise to the microblazar one-sided morphology of \lsi{}
 \citep{Massietal2013, Massi2014}.

Is the variation of the ejection angle random or is there a
periodicity? The motion of a jet can be revealed by the path traced by
its core in consecutive images. The astrometry result in
\citet{Massi2012} is that the core describes a periodic path with a
precession period of 27--28~days \citep{Massi2012}. This determination
has been confirmed by the timing analysis of 36.8~years of radio data
\citep[][and references therein]{Massi2016}. The results of the timing
analysis are two dominant features at $P_1 = 26.496 \pm 0.013$~d (the
orbital period) and $P_2 = 26.935 \pm 0.013$~d, and the period\ $P_2$
is clearly fully consistent with the period of 27--28~days that was
determined by VLBA astrometry.

\subsection{Beating between orbit ($P_1$) and precession ($P_2$)}
\label{sect:beating}

The timing analysis by \citet{Massi2016} shows, as a minor feature
with respect to $P_1$ and $P_2$, the long-term period $P_{\rm long} =
1628 \pm 48$~d. The same feature appears as well in the timing
analysis of model data of a precessing ($P_2$) conical jet refilled
periodically (with $P_1$) with relativistic electrons (Figs~4\,a and
4\,b in \citealt{Massi2016}). Since $P_{\rm long}$ was not given as
input in the model, this implies that $P_{\rm long}$ is the result of
$P_1$ and $P_2$. In fact, the beating of $\nu_1$ (1/$P_1$) and $\nu_2$
(1/$P_2$) gives rise to a modulation with period $P_{\rm beat} =
{1\over \nu_1 -\nu_2} = 1626 \pm 68$~d, i.e. the long-term
modulation.

Along with the identity of $P_{\rm long}$ with $P_{\rm beat}$, an
additional important point is that neither $\nu_1$ nor $\nu_2$ but
rather their average, $P_{\rm average} = {2\over \nu_1 + \nu_2}$, is
modulated in a beating. The average of $P_1 = 26.496 \pm 0.013$~d and
$P_2 = 26.935 \pm 0.013$~d is  26.71~d, and indeed 
the observed periodicity of the radio outbursts in \lsi,{}  is
$26.704 \pm 0.004$~d \citep{Ray1997, Massi2013, Jaron2013}.

The fact that the periodicity of the outburst is $P_{\rm average}$
explains the well-known problem of the  orbital shift of the
outburst or timing residuals problem \citep[and references
   therein]{paredes90, Gregory1999}. Orbital shifts or timing
residuals occur because the outburst does not have orbital
periodicity, i.e. $P_1$, but is in fact periodic with $P_{\rm
  average}$. \citet{Massi2013} have shown mathematically that the
timing residuals correspond to the difference between $P_1$ and
$P_{\rm average}$. When the outbursts are correctly folded with
$P_{\rm average}$, they cluster perfectly without any shift
\citep[Fig.~1 in][]{Massi2016}.

Direct determination of $P_1$ and $P_2$ in timing analysis or the
determination of long-term modulation and orbital shift of data folded
with $P_1$ are, therefore, equivalent indicators that the intensity
variations are induced by the precessing jet. The first method finds
the two physical periods directly, and the second method deals with
the result of the beating. In addition to radio data, the periods
$P_1$ and $P_2$ were also directly determined in the Lomb-Scargle
spectrum of data at apastron for \fermilat{} data \citep{Jaron2014}
and X-ray data \citep{D'Ai2016}, whereas the long-term modulation
\citep{Zamanov2013} and orbital shift \citep{Paredes-Fortuny2015} were
found in the equivalent width EW of the H$\alpha$ emission line.

\subsection{The disk of the Be star}
\label{sect:Bedisk}

\begin{figure}
  \includegraphics{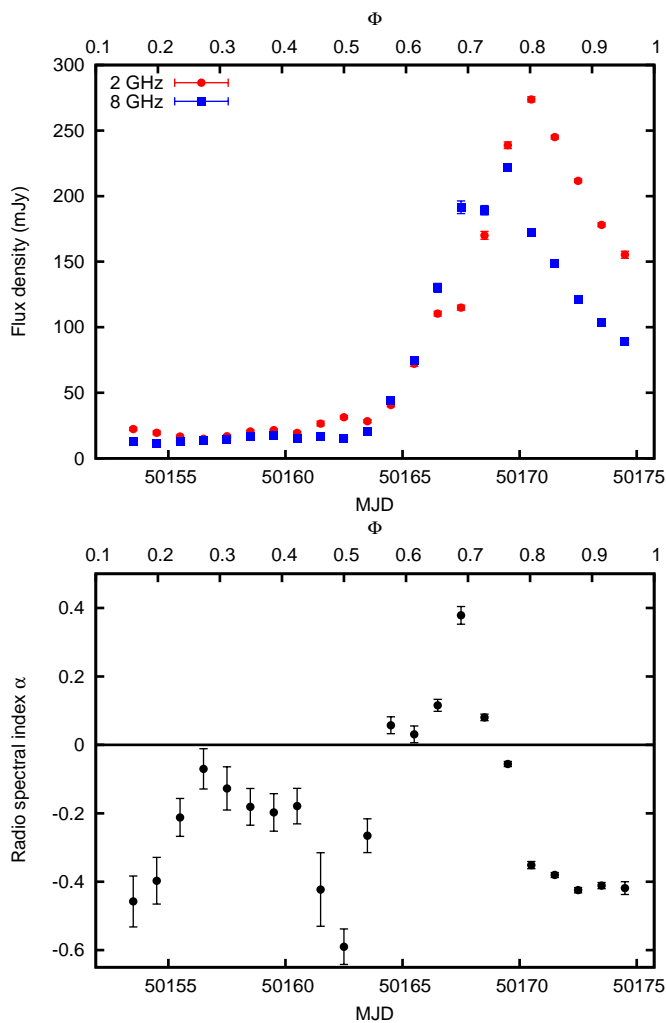}
  \caption{
    Top: Radio light curve at 2 (red circles) and 8 GHz (blue squares)
    observed by the GBI from MJD~50154.6 until 50175.7 (orbital phase
    $\Phi = 0.1-1.0$), averaged over one day.
    Bottom: Radio spectral index~$\alpha$ of the radio fluxes shown in
    the top panel clearly follows a two-peaked trend. One peak is
    during the large apastron radio outburst, which features a flat or
    inverted ($\alpha \geq 0$) spectrum during its rise.
    The other peak of
    the spectral index occurs at $\Phi = 0.25$, i.e. at periastron,
    revealing that the spectrum becomes flattish ($\alpha \approx 0$)
    during this orbital phase as well \citep{Massi2009}.
  }
  \label{fig:radio_alpha}
\end{figure}

Before the discovery of the beating between $P_1$ and $P_2$, the
assumptions for the long-term modulation were variations of the Be
wind \citep{Gregory2002b} with the further constraint of Be wind
long-term variations influencing only the apastron and not the whole
orbit because  long-term variations in the emission only occur towards
apastron \citep{Ackermann2013}. Whereas in the previous section we
have seen the evidence for beating, in this section we report on
observations probing the stability over the last 36.8~years of the Be
disk.

Cycles in varying Be stars change in length and disappear after 2--3
cycles, following the well-studied case of the binary system $\zeta$
Tau \citep[][and references therein]{Stefl2009}. In this context,
radio variations induced by variations in the Be disk are predicted to
be rather unstable. A powerful method for studying the stability of a
periodic signal over time is its autocorrelation. If a period is
stable the correlation coefficient should appear as an oscillatory
sequence with peaks at multiples of that period. The analyses of
36.8~years of radio data \citep{Massi2016}, corresponding to eight
cycles of the long-term modulation, have definitely shown a very
stable pattern for the radio data. The correlation coefficient of the
observations shows a regular oscillation at $P = 1626$~d (their
Fig.~5) as does the correlation coefficient of the model data of a
precessing jet.

A stable $P_{\rm long}$ is what is expected for a beating of periods
$P_1$ (orbital period) and $P_2$ (precession) without significant
variations during the eight examined cycles \citep{Massi2016}. A
stability in $P_1$ implies the same physical conditions in the
accretion. This means that the decretion disk of the Be star was
stable during the last 36.8~years. This is consistent with the
behaviour of Be stars. Over the last 100~years, the Be star $\zeta$
Tau, whose emission has never disappeared completely, has gone through
active stages, characterized by pronounced long-term variations, and
quiet stages \citep{Harmanec1984}. For a period of 30~years, from 1920
to about 1950, there were no variations.

The implications for the accretion along the orbit of \lsi{} is that
once the Be wind parameters such as as density and velocity (see
Sect.~\ref{sect:introduction}) are relatively stable, the two
accretion peaks from the compact object should always occur at the
same two orbital phases. Is there any evidence for these predicted
peaks at $P_1$? \fermilat{} data over the large interval of 7.6~years,
when folded on the orbital period (Fig.~\ref{fig:folded}) hint at two
peaks. Indeed, timing analysis of this data results in a periodicity
$P_1$ at periastron (Fig.~\ref{fig:timing}\,a) \textit{and} towards
apastron (Fig.~\ref{fig:timing}\,c), as already shown in
\citet{Jaron2014} with a smaller database. If the accretion peaks give
rise to two ejections, then these two ejections, in the context of
microquasars, should produce a flat spectrum twice along the
orbit. Figure~2 shows the flux densities, $S_1$ and $S_2$, at two
frequencies and their radio spectral index, $\alpha =
{\log(S_1/S_2)\over \log(\nu_1/\nu_2)}$, along one orbital
cycle. During the onset of the radio outburst $\alpha \ge 0$ as for
microquasars, and attains a higher value at orbital phase 0.7
(i.e.\ apastron). In addition to that, around phase 0.2--0.3, i.e. at
periastron, even if the flux density is very low, $\alpha$ shows a
clear evolution, rises and reaches nearly zero at periastron, remains
flattish, and then decays at the low value of $-0.6$. This trend of
$\alpha$ is discussed for a radio database of years in
\citet{Massi2009}. As is shown in this paper, electrons suffer strong
IC losses at periastron, the result of which is a short
jet giving rise to a little flux. Nevertheless, the little flux does
not lose its spectral property and the ratio of flux density at two
frequencies (i.e. $\alpha$) keeps the main characteristics as for the
strong emission at apastron. Finally, optical observations by
\citet{Mendelson1989} showed two variations in orbital phase: one
variation at periastron and another smaller variation at
apastron. This optical emission could come from the hot accretion flow
(optical synchrotron emission) as studies of X-ray binaries suggest 
\citep{Poutanen2014}.

\section{Methods}
\label{sect:methods}

In the following we extend the physical model of a precessing ($P_2$)
jet periodically ($P_1$) refilled with relativistic electrons
developed in \citet{Massi2014}. Here the jet is refilled in two
different parts of the orbit (Sect~\ref{sect:rel_el}) and not only at
apastron as in \citet{Massi2014}. The synchrotron emission of such a
jet is calculated as in \citet{Massi2014}; in addition we calculate
IC radiation here. Seed photons are stellar photons (external inverse
Compton; EIC) (Sect~\ref{sect:stellar_seed_photons}) and the photons
emitted by synchrotron radiation (synchrotron self Compton; SSC),
i.e. jet photons (Sect.~\ref{sect:jet_seed_photons}). The total IC
emission from the model is calculated
(Sect.~\ref{sect:ic_emission}). Electron energy losses are taken into
account and the length of the jet is determined (in
Sect.~\ref{sect:accelerated_electron_survival}).

\subsection {Relativistic electron distribution}
\label{sect:rel_el}
 
Accretion theory predicts two maxima for the eccentric orbit of \lsi{}
\citep{Taylor1992, Marti1995, Bosch-Ramon2006, Romero2007}. One
maximum is around periastron and the second is towards apastron. We
indicate the two relativistic electron distributions as $N_{\rm I}$
(periastron) and $N_{\rm II}$ (towards apastron).

Quantities of this section include functions of the orbital
phase\ $\Phi$, or of the long-term modulation\ $\Theta$, defined
as \begin{eqnarray}
  \Phi & = & \frac{t - t_0}{P_1} - {\rm int}\left(\frac{t -
    t_0}{P_1}\right),\\
  \Theta & = & \frac{t - t_0}{P_{\rm beat}} - {\rm int}\left(\frac{t -
    t_0}{P_{\rm beat}}\right),
\end{eqnarray}
where $t_0 = \unit[43366.275]{MJD}$ and ${\rm int}(x)$ takes the
integer part of $x$.

In \citet{Massi2014}, the radio emission was reproduced with a
relativistic electron distribution\ $N$ ejected around apastron which,
once expressed in ${\rm electrons} \over {\rm cm^{3} d\gamma}$, is
written as
\begin{eqnarray}
  \label{eq:NII}
  N_{\rm II} & = 
   K_{\rm II}(\Phi)l^{-a_3} \gamma^{-p} ,
\end{eqnarray}
with  $p = 1.8$ and $a_3 = 2(2+p)/3$ as derived by \citet{Kaiser2006}
for the adiabatic jet to take into account losses due to jet
expansion.

In the present work we add another component to the above electron
distribution, with a maximum around periastron, as follows:
\begin{equation}
  N_{\rm I}  = K_{\rm I}(\Phi)l^{-a_3}\gamma^{-p}.
\end{equation}
This new electron component is injected and expands in the same
conical magnetic structure defined by the model of \citet{Massi2014}.

\begin{figure}
  \includegraphics{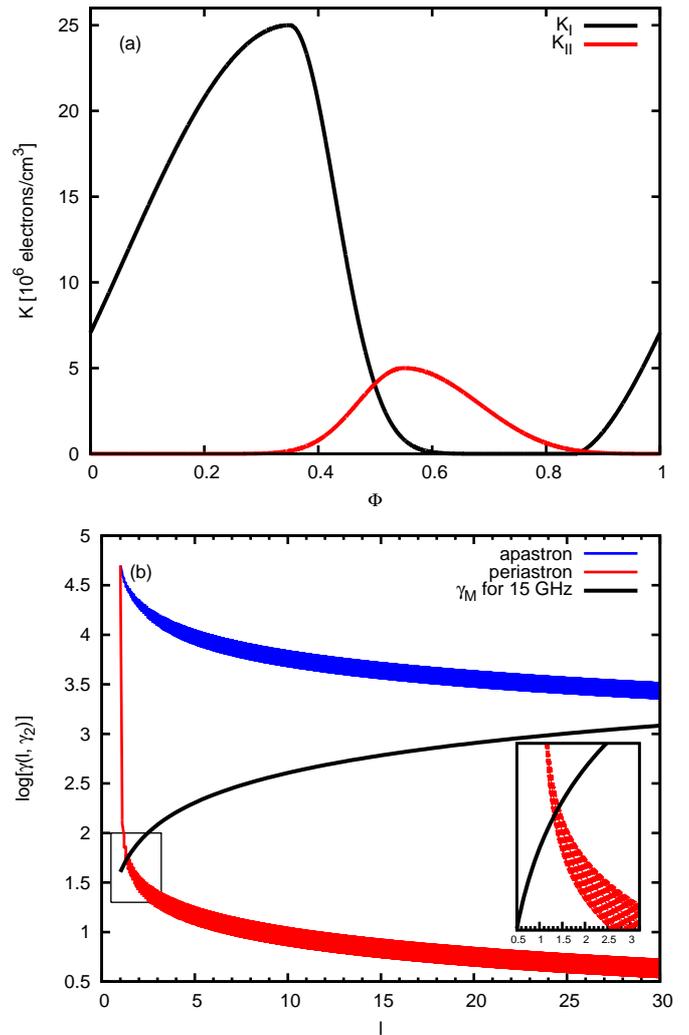}
  \caption{
    (a) Orbital modulation of the electron density assumed in the
    model, as defined in Sect.~3.1. The two ejections are given at
    $\Phi = 0.35$ and  at $\Phi = 0.58$.
    (b) The maximum value of the Lorentz factor $\gamma$ of the
    electrons as a function of the position $l$ along the jet for the
    two ejections. The black curve is a plot of the minimum $\gamma$
    (Eq.~\ref{eq:gammamin}) necessary for the production of radio
    emission at 15~GHz for an initial magnetic field $B_0 = 5$~G.
    The intersection point of the black curve with the red and
    blue lines defines the length of the jet at periastron and that of
    the jet towards apastron, respectively.
  }
  \label{fig:losses}
\end{figure}

We assume that the accelerated electrons powered by accretion are
injected both at periastron and apastron in similar ways and that the
initial (i.e.\ at time $t = t_{\rm min}$ in $l = 1$) electron energy
range is the same for the two distributions,
\begin{equation}
  \gamma_1 \leq \gamma(t_{\rm min} ) \leq \gamma_2.
\end{equation}
However, as we see in Sect.~\ref{sect:accelerated_electron_survival},
the value of $\gamma$ evolves in a different way at periastron than at
apastron owing to the different weight of radiative losses.

In our code each ejection of electrons is a sine function (of orbital
phase) with different exponents ($n$, $ex$) for onset and decay. At
each observing time the orbital phase and contribution of the sine
functions for that phase are calculated. The results of the fit are
position, amplitude, and exponents for both ejection functions (see
Fig.~3 and Table~1).

For the specific case shown in Fig.~\ref{fig:losses}, we computed the
mean amount of energy injected per second in the form of relativistic
electrons at the jet basis, $l = 1$. Taking a mean over the orbital
phase of the electron density normalization factor, $K_{\rm mean} =
\int^1_0 [K_I+K_{II} ]~d\Phi$, we obtain
\begin{equation}
  L_{\rm rel} = 2\left(\pi r_o^2c\right)~K_{\rm mean}
  \left(mc^2\right)^{2-p} \int^{\gamma_{2}}_{\gamma_{1}} \gamma^{1-p}
  d\gamma \simeq \unit[8\,10^{35}]{ergs/sec}
\end{equation}
for the two jets. Here $r_0 = x_0tg\xi$ is the initial jet radius and
$x_0$ is a normalization length for the jet, which starts at $x = x_0$
and extends to $x = x_0L$ with an opening angle $\xi$.

The orbital dependence of the injected energy follows the same
modulation shown in Fig.~\ref{fig:losses} for the relativistic
electron density with a maximum around $\Phi \sim 0.35$ for which we
have $[ L_{\rm rel}]_{MAX }\simeq  1.8~10^{36}$ ergs/sec.

\subsection{Stellar seed photons}
\label{sect:stellar_seed_photons}

In order to take into account IC scattering of B0 star photons by
relativistic electrons in the jet, it is necessary to compute the
distance between the central B0 star and the generic position,
  $x_0l$, along the axis of the jet anchored to the orbiting compact
object. With $\psi$ being the opening angle of the precession cone,
under the assumption that the jet precession axis is perpendicular to
the orbital plane, we can compute
\begin{eqnarray}
  && d_{\perp} =  x_0l\cos\psi,\\
  && d_{\rm plane} =  \sqrt{r(\Phi)^2 + (x_0l\sin\psi)^2 -
    2r(x_0l\sin\psi)\cos\Omega},\\
  && d_{\rm jet-star}(l, t) =  \sqrt{d_{\rm plane}^2 +
    d_{\perp}^2},
\end{eqnarray}
where $\Omega$ (as in \citealt{Massi2014}) introduces the
long-term modulation phase dependence and $r(\Phi)$ is the distance
from the Be star to the point of the orbit corresponding to phase
$\Phi$. A sketch of the geometry is shown in Fig.~\ref{fig:timing}.

Assuming for the star a blackbody emission with temperature~$T_*$,
the photon density at the position $x_0l$, along the jet is
\begin{equation}
  \rho_{\rm star}(l, \Phi) = \unit[{4\pi B_{\nu}(T_*) \over ch^2\nu}
    {\pi R_*^2 \over 4 \pi d_{\rm
        jet-star}^2}]{\left[photons~cm^{-3}erg^{-1}\right]}.
  \label{eq:star_den}
\end{equation}
It is evident that this density value can only be affected by the
different position along the orbit of the compact object, and hence of
its associated jet, if the values of $r(\Phi)$ and $x_0$ are
comparable. Hence, the orbital periodicity in the IC reprocessing of
stellar photons show up only if relativistic electrons are present at
distances from the orbital plane of the order of the ellipse's
semi-major axis, $a$, which in our case implies distances $\simeq
7~10^{12}$~cm \citep{Massi2012}. This fact implies that the jet scale
dimensions, $x_0$, must be of this order of magnitude or even
smaller.

The energy of IC reprocessed photons is linked to both electron and
seed photon energies since the reprocessed photon energy, $\epsilon$,
fullfils the condition $\epsilon \sim \gamma^2 h\nu_{\rm seed}$. From
the above Eq.~(\ref{eq:star_den}) it is easy to derive that the
maximum of the stellar photon number density is located at $\nu_{\rm
  peak}$ such that $h\nu_{\rm peak}/ kT_* \simeq 1.6$. The photon
density rapidly decreases for larger frequencies, while the decrease
is slower for lower frequencies so that for $h\nu/kT \simeq 0.1$ the
photon density is around 1/7 of its maximum value. Hence, in our case
with $T_* = 28000$~K, for $\nu_{\rm peak}/16 \leq \nu_{\rm seed} \leq
\nu_{\rm peak}$, relativistic electrons with values of the Lorentz
factor up to $\gamma \sim 7~10^4$ can ensure that the reprocessed
photons has energy of the order of 1~GeV, i.e. in the observational
range chosen in this paper (see Sect.~\ref{sect:observations}).

\subsection{Jet seed photons}
\label{sect:jet_seed_photons}

The jet photon density, owing to synchrotron emission inside the jet,
can be computed as a function of the distance $l$ along the jet axis,
\begin{equation}
  \rho_{\rm jet}(l) = \unit[\frac{4\pi
      I_{\nu}(l)}{ch^2\nu}]{\left[photons\,cm^{-3}\,erg^{-1}\right]}.
\end{equation}
In this expression, $I_{\nu }$ is the locally produced synchrotron
emission in the optically thin limit
\begin{equation}
  I_{\nu }(l) = \int J_{\nu} x_0 {\rm d}l,
\end{equation}
and the notation is the same as in \citet{Massi2014}. The above photon
density is monotonically decreasing as $\nu^{-(p+1)/2}$ with a maximum
in the radio range where emission becomes optically thick.

Different energy ranges in synchrotron photon spectrum contribute to
different energy ranges in the reprocessed SSC emission and the only
condition we  require is that reprocessed photons with energy in
the $0.1-3$~GeV band can be produced. \citet{Ginzburg1965} show that
synchrotron emission is centred on a peak spectral frequency $\nu$
such that
\begin{equation}
  \nu_{\rm sync.} = 1.8~10^6 \gamma^2 B.
  \label{eq:syn}
\end{equation}
Hence, our condition reads
\begin{equation}
  \gamma^2 h\nu_{\rm sync} = 1.8~10^6\gamma^4 h B \sim \unit[1]{GeV}.
\end{equation}
This condition is fulfilled if accelerated electrons exist with values
of the Lorentz factor up to $\gamma \sim 2~10^4$ with $B_0 \sim
1$~gauss. This condition is not as strong as the previous condition found
for stellar seed photons.

\subsection{Inverse Compton emission}
\label{sect:ic_emission}

In order to compare our theoretical model for GeV emission to
available data we derive the quantity
\begin{equation}
\label{eq:flux}
  F_{\rm 0.1-3\,GeV} = \unit[{1 \over 4 \pi D^2} \intop^{\rm 3\,GeV}_{\rm
      0.1\,GeV}{E_{\rm IC} \over \epsilon} {\rm
      d}\epsilon]{\left[counts\over cm^2 sec\right]},
\end{equation}
where $D$ is the distance, $\epsilon = h\nu$ is the photon energy, and
$E_{\rm IC}$ is the total scattered power per energy (in units of
erg\,sec$^{-1}$\,erg$^{-1}$). This can be computed following the
derivation by \citet{Rybicki1986} for the isotropic case,
\begin{equation}
  E_{\rm IC} =\sum_i \intop_V {A~Q_{\rm i}\over l^{a_3}}\int
  {\epsilon \over \epsilon_{\rm S}}\rho_i(\epsilon_{\rm S})
  \intop^{\gamma_M}_{\gamma_m}\gamma^{-p-2} f\left({\epsilon \over 4
    \gamma^2 \epsilon_{\rm S}}\right){\rm d}\gamma{\rm
    d}\epsilon_S{\rm d}V,
    \label{eq:emis}
\end{equation}
where the overall emission is composed of the sum of two different
contributions: SSC from jet photons ($i = 1,\, \rho_1(\epsilon_{S}) =
\rho_{\rm jet}$) and EIC from star photons ($i = 2,\,
\rho_2(\epsilon_{S}) = \rho_{\rm star}$).

In the above expression $\epsilon_{\rm S} = h\nu_{\rm S}$ is the seed
photon energy, $V$ is the volume over which to integrate the IC
emission, i.e. the jet and other quantities are defined as follows:
\begin{equation}
  f \left( {\epsilon \over 4 \gamma^2
    \epsilon_{\rm S }}\right) = {2 \over 3} \left(1-{\epsilon \over 4
    \gamma^2 
    \epsilon_{\rm S}}\right)~~~~{\rm for} ~~~~~~~0\leq {\epsilon \over
    4 \gamma^2 
    \epsilon_{\rm S }}\leq 1,
\end{equation}
and zero otherwise,
\begin{equation}
  Q_i = K_{\rm I} \times DB_{i}(\beta_{\rm I})+K_{\rm II}\times
  DB_i(\beta_{\rm II})
\end{equation}
\begin{equation}
  A  = {3 \over 4} c \sigma_{\rm T},\\
\end{equation}
with the Thompson cross-section\ $\sigma_{\rm T}$. The above
expressions for plasma emission holds in the frame in which the plasma
bulk motion is zero; the Doppler boosting  (DB) term, $DB_{\rm i}$, takes
into account that the jet emitting plasma is moving with a speed\ $v$
along the jet axis. Its general expression is written as
\begin{eqnarray}
  DB_{\rm i} & = & \left[{1\over \Gamma(1 \pm
      \beta\cos\eta)}\right]^{ex_i},\\
  \Gamma & = & {1 \over \sqrt{1-\beta^2}}, \nonumber
\end{eqnarray}
with $\beta = v/c$, and the possibility for a different injection
velocity for each distribution is allowed, i.e. $\beta = \beta_{\rm
  I}$ for periastron distribution and $\beta = \beta_{\rm II}$ for
apastron distribution. The parameter $\eta$ is the angle between the
observer's line of sight and the jet axis (see \citealt{Massi2014} for
details). The DB exponent\ $ex_i$ for IC radiation
produced by stellar seed photons is different from that for
synchrotron jet photons and, in particular, we have  $ex_1 = 3 -
\alpha$ (see \citealt{Ghisellini1985}, where the spectral index
$\alpha$ is defined with an opposite sign with respect to our
definition) and $ex_2 = 2 + p$ \citep{Kaufman2002}.

The above expression (\ref{eq:emis}) holds in the Thomson limit
approximation, i.e. for $\gamma\nu_{S}h < mc^2$. This limit is not
overpassed for the range of energies of \fermilat{} data analysed in
this paper and with the above described seed photon energy densities,
if we set the free parameter $\gamma_2 \leq 7~10^4$.

The integral over the seed spectrum, i.e. on $\epsilon_{\rm S}$, is
extended to all seed photons that can contribute to GeV
emission. Since the function\ $f\left({\epsilon \over 4 \gamma^2
  \epsilon_{\rm S}}\right)$ is different from zero only in the
interval $[0, 1]$, this condition shows that, given an electron
distribution in a specific $\gamma$ range, and a seed photon of energy
$\epsilon_{\rm S}$, IC emission is different from zero only for
energies, $\epsilon$, such that $0 \leq \epsilon \leq 4\gamma^2
\epsilon_{\rm S}$.

The integration interval in ${\rm d} \gamma$ depends on the distance
$l$ along the jet axis since electrons lose energy while proceeding
along the jet. Therefore, the integration range is $\gamma_M(l = 1) =
\gamma_2$ and $\gamma_m(l = 1) = \gamma_1$ at the beginning of the
jet, while for a generic distance we have $$\gamma_M(l) = \gamma (l,
\gamma_2)~~~~~~\gamma_m(l) = \gamma (l, \gamma_1),$$ where these
functions are specified by Eq.~(\ref{ga_losses}) in the next section.

\subsection{Accelerated electron survival}
\label{sect:accelerated_electron_survival}

As outlined in various parts of this section, relativistic electrons
lose energy with time, i.e. on their way along the jet, owing to
different types of losses and hence their $\gamma$ values
decrease. The standard way to take  electron energy losses into
account requires the solution of the following equation (see
e.g.\ \citealt{Longair1994}):
\begin{equation}
  \label{eq:g_losses}
        {{ \partial  \gamma  \over \partial t} = -
          \gamma {
            2~a_1 \over  3~ t}- W~\gamma^2 U_B-
          W~\gamma^2 U_{jet}- W~\gamma^2 U_{star} .   
        }
\end{equation}
The first term on the right-hand side takes adiabatic losses into
account, as derived by \citet{Kaiser2006} for the specific jet
configuration we are using in this paper. The following terms account
for synchrotron and IC losses due to  electron interactions with synchrotron
jet photons (SSC) and star photons (EIC). In the
above expression $W = 4 \sigma_T /(3m_e c) = \unit[3.2~10^{-8}]{sec~cm
  \over g}$, the term
\begin{equation}
  U_B(l) =  \unit[{B_0^2 \over 8\pi}l^{-a_2}]{\left[erg\,cm^
      {-3}\right]}
\end{equation} 
defines the magnetic energy density, while the two terms
\begin{equation} 
  U_{\rm jet}(l, \Phi) = \unit[{4 \pi \over c}\int I_{\nu}{\rm
      d}\nu]{\left[erg\,cm^{-3}\right]}
\end{equation}
\begin{equation}
  U_{\rm star}(l, \Phi, \Theta) = \unit[\int{B_{\nu}(T_*) \over c}{\pi
      R_*^2 \over d_{[\rm jet-star]}^2}~{\rm d}\nu]{\left[erg\,cm^
      {-3}\right]}
\end{equation} 
describe synchrotron photon density and star photon density inside the
jet.

The most likely magnetic configuration in a jet is a helical magnetic
field configuration with both the parallel and perpendicular
components. Following Kaiser's model we can reproduce $B\sim
B_{\parallel}$ by setting $a_2 = 2$ and $B \sim B_{\perp}$ with $a_2 =
1$. In our framework, as in \citet{Massi2014}, we set for our model
$a_1 = 1$, corresponding to a jet with a constant opening angle, and
$a_2 = 2$, corresponding to a parallel magnetic field, since we found
comparable solutions for the two configurations of $B$ as in
\citet{Massi2014}.

Equation~(\ref{eq:g_losses}) is a first order differential equation of
the Bernoulli type, which can be easily integrated between $t$ and
$t_{min}$ to give
\begin{equation}
  \label{eq:Kaiser_like}
        { \gamma(t) = {
            {\it g}t^{-2/3}
            \over
            t_{\rm min}^{-2/3} + W 
            \it g \int^t_{t_{min}}t^{-2/3}\left[U_B + U_{\rm
                jet} + U_{\rm star}\right]  ~dt
        }}
        ,
\end{equation}
where ${\it g}=\gamma(t_{\rm min})$. Setting $t_{\rm min} =
x_0/(\Gamma\beta c)$ for electrons starting at $l = 1$, i.e. at the
jet basis, $x_0$, and $t = {x_0 l/(\Gamma\beta c)}$ as the time at
which electrons attain the generic position $lx_0$ along the jet axis,
Eq.~(\ref{eq:Kaiser_like}) can be rewritten in terms of the spatial
coordinate along the jet axis as
\begin{equation}
  \gamma(l, g) =
  \frac{
    g l^{-2/3}
  }{
    1 + W~\frac{
      x_0
    }{
      \Gamma\beta c
    }
    g\left\{\int^l_{1}l^{-2/3}\left[U_B + U_{\rm
        jet} + U_{\rm star}\right]{\rm d}l\right\}
  }.
  \label{ga_losses}
\end{equation}
This expression shows how $\gamma$ evolves along the jet when subject
to adiabatic, synchrotron, and IC losses.

Following Eq.~(\ref{ga_losses}) each initial Lorentz factor value,
${\it g}$, decreases along the jet so that the entire initial
distribution shifts to lower energies. In particular, choosing
$\gamma(t_{\rm min}) = {\it g} = \gamma_2$, we can determine how the
upper limit of the distribution evolves along the jet,
i.e. $\gamma_M(l) = \gamma (l, \gamma_2)$.  This evolution 
is shown in Fig.~\ref{fig:losses}\,b. In our present
model we account for all types of energy losses described in this
section through the use of a different upper limit cut-off of the
electron distribution for each position $l$ along the jet axis; this
is the same method used by \citet{Kaiser2006} only for adiabatic and
synchrotron losses.  As stated in Sect. 3.1,  losses due to adiabatic jet expansion
have been taken into account  also in  the electron density
distribution [see Eq.(\ref{eq:NII})].
 
Using the above cited Eq.~(\ref{eq:syn}), we can derive the jet
length as the position along the jet axis above which electron
energies are too low to emit the observed radio flux. In fact, for $B
= B_0 l^{-2}$ and with a magnetic field value at the jet basis of $B_0
= 5.1$~G (as results from our present model), it is possible to have
synchrotron radiation at $\nu = 15$\,GHz (OVRO observations) only if
the electron distribution extends up to
\begin{equation}
  \gamma_M(l) > \left[\frac{\nu~l^2}{1.8\,10^6B_0}\right]^{1/2}
  \approx 40~l.
  \label{eq:gammamin}
\end{equation}
Condition (\ref{eq:gammamin}) is drawn as a black line in
Fig.~\ref{fig:losses}~b; its intersection with the red (blue) line
describing the $\gamma$ decrease induced by losses at periastron
(apastron) defines the jet length $L \sim 1.2$ ($L \sim 25$).
At apastron it is evident that the jet length ($L \sim 25$), and hence
the radio emission, is not limited by radiative losses.

\section{Observations and data reduction}
\label{sect:observations}

We compare the model output data to observational data at radio and
GeV wavelengths. The databases that we use are the OVRO monitoring at
15~GHz ranging from MJD~54684 to MJD~56794 (2008~Aug~6 to
2014~May~17), and GeV $\gamma$-ray data from the \fermilat{} in the
energy range 0.1--3.0~GeV from MJD~54682 to MJD~57450 (2008~Aug~4 to
2016~March~3).

The calibration of radio OVRO data is described in \citet{Massi2015}.

We used version v10r0p5 of the Fermi ScienceTools\footnote{available
  from
  \url{http://fermi.gsfc.nasa.gov/ssc/data/analysis/software/}}for the
analysis of \fermilat{} data. We used the instrument response function
P8R2\_SOURCE\_V6 and the corresponding model gll\_iem\_v06.fits for
the Galactic diffuse emission and the template
iso\_P8R2\_SOURCE\_V6\_v06.txt. Model files were created automatically
with the script make3FGLxml.py\footnote{available from
  \url{http://fermi.gsfc.nasa.gov/ssc/data/analysis/user/}} from the
third \textit{Fermi}-LAT point source catalogue
\citep{FermiLAT2015}. The spectral shape of \lsi{} in the GeV regime
is a power law with an exponential cut-off at 4--6~GeV
\citep{Abdo2009, Hadasch2012}. Here we restrict our analysis to the
power law part of the GeV emission by fitting the source with
\begin{equation}
  \frac{{\rm d}n}{{\rm d}E} = n_0\left(\frac{E}{E_0}\right)^{-\alpha +
    \beta\log\left(E/E_{\rm b}\right)}~~\left[{\rm counts \over cm^2
      sec~ dE}\right],
\end{equation}
where all parameters are left free for the fit and including data in the
energy range $E = \unit[0.1 - 3]{GeV}$. All other sources within a
radius of $10^{\circ}$ and the Galactic diffuse emission were left
free for the fit. All sources between $10-15^{\circ}$ were fixed to
their catalogue values. The light curves were computed by performing
this fit for every time bin of width one day for \fermilat{} data from
2008~August~8 (MJD~54684) till 2016~March~3 (MJD~57450). On average,
the test statistic for \lsi{} was 40, which corresponds to a detection
of the source at the $6.3\sigma$ level on average in each time bin.

The search for periodicities in the light curves is carried out with
the UK Starlink package, implementing the Lomb-Scargle algorithm
\citep{Lomb1976, Scargle1982}. The procedure is the same as outlined
in \citet{Massi2013}.

\subsection{Consistency with previous results}

\begin{figure}
  \includegraphics{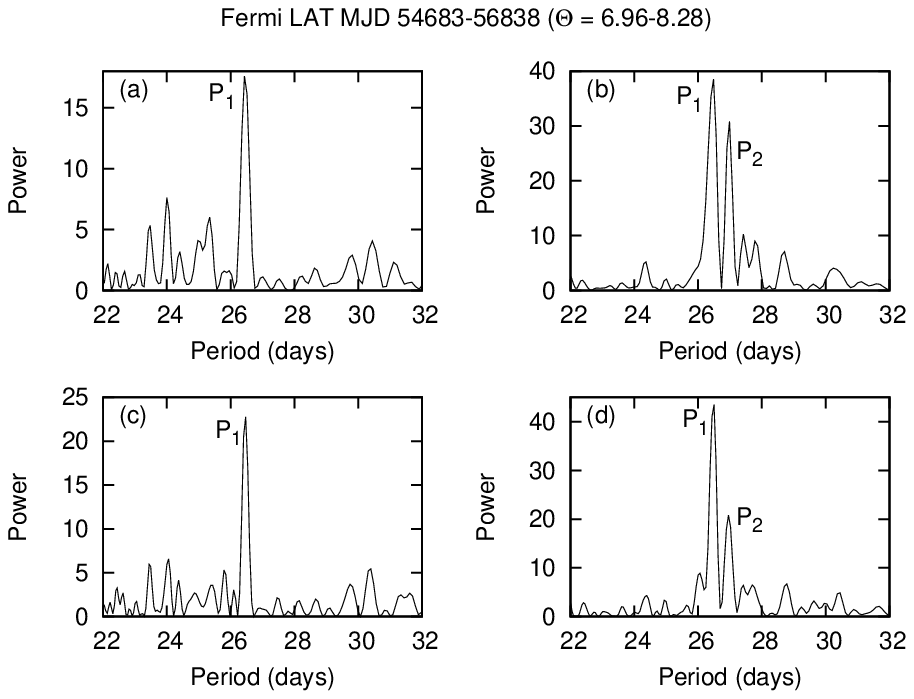}
  \caption{
    Timing analysis of Pass~8 \fermilat{} data, energy range
    0.1--3.0\,GeV. (a, b) Time interval MJD~54683--56838 (same epoch as used
    in \citealt{Jaron2014}, where they used Pass~7 data and the energy
    range 0.1--300\,GeV). The result of the timing analysis is
    confirmed (compare Fig.~3 in \citealt{Jaron2014}):
    (a) periastron, only $P_1$; and    (b) apastron, $P_1$ and $P_2$.
    (c, d) Time interval is 54683--57450:
    (c) periastron, only $P_1$; and
    (d) apastron, $P_1$ and $P_2$, but with less power than in (b).
  }
  \label{fig:compJM14}
\end{figure}

\citet{Jaron2014} found that the GeV data from \lsi{} observed by the
\fermilat{} are modulated by $P_1$ and $P_2$ in the orbital phase
interval\ $\Phi = 0.5 - 1.0$ (apastron), but by only $P_1$ in $\Phi =
0.0 - 0.5$ (periastron) (see their Fig.~3). Their analysis was
performed with Pass~7 \fermilat{} data, using the energy range
$0.1-300$\,GeV, and for the time interval MJD~54683--56838. We use
Pass~8 \fermilat{} data and restrict the energy range to $0.1 -
3.0$\,GeV. It is therefore important to verify how the data processed
with the new method compare to the previous data. We apply the timing
analysis to the data processed with the new method as in
\citet{Jaron2014}. The result is shown in the top panel of
Fig.~\ref{fig:compJM14}. Since we use a time bin of one~day for the
likelihood analysis, Figs 3\,e and h in \citet{Jaron2014} are the
figures to use for comparison. Clearly the result of \citet{Jaron2014}
is confirmed here; in our Fig.~\ref{fig:compJM14}\,a the only
significant feature is the peak at the orbital period\ $P_1$, whereas
in Fig.~\ref{fig:compJM14}\,b there is not only a peak at $P_1$ but
also a highly significant peak at $P_2$. This result shows that the
timing characteristic reported by \citet{Jaron2014} is  a feature from
the emission in the range 0.1--3\,GeV.

\subsection{Influence of the $\Theta$ interval on $P_2$}

Applying the same timing analysis to all available \fermilat{} data
ranging from MJD~54682 to MJD~57450 ($\Theta = 6.8 - 8.8$) we obtain
the result presented in the bottom panel of
Fig.~\ref{fig:compJM14}. In Fig.~\ref{fig:compJM14}\,c, which shows
the Lomb-Scargle periodogram for the periastron data, the orbital
period\ $P_1$ has increased its power and stands out a bit more
significantly over the noise when compared to
Fig.~\ref{fig:compJM14}\,a. However, the $P_2$ feature in
Fig.~\ref{fig:compJM14}\,d has decreased its power and its relative
importance with respect to $P_1$ drops to about 1/2, when it was 3/4
for the narrower time range shown in Fig.~\ref{fig:compJM14}\,b.

\begin{figure}
  \includegraphics{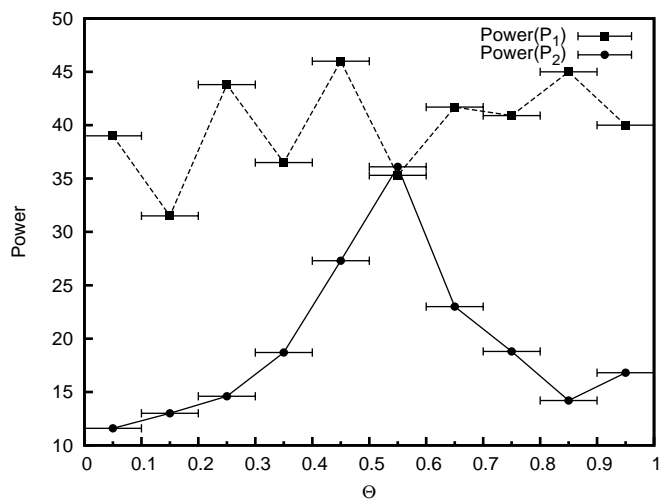}
  \caption{
    Test on the $\Theta$ interval that corrupts the timing
    analysis. In this plot the powers of $P_1$ and $P_2$ resulting
    from Lomb-Scargle timing analysis are plotted as a function of the
    $\Theta$ interval removed from the data, where interval is
    indicated with horizontal error bars. While the power of $P_1$
    is only affected very little and in a random way, the power of
    $P_2$ shows a systematic trend peaking at $\Theta = 0.5-0.6$.
  }
  \label{fig:searchP2}
\end{figure}

\begin{figure}
  \includegraphics{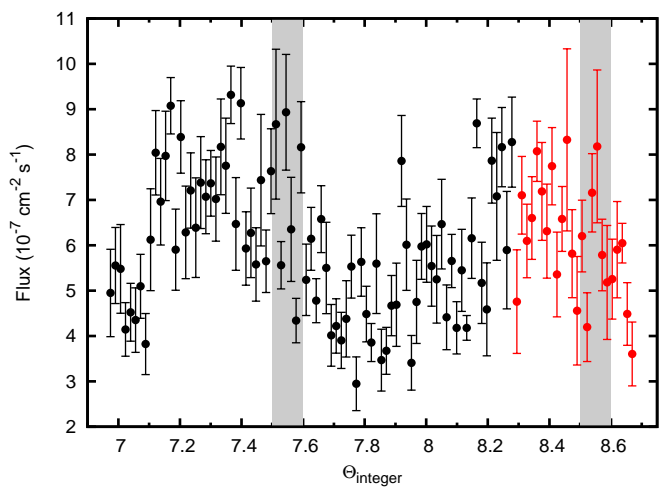}
  \caption{
    All \fermilat{} data from the orbital phase interval $\Phi =
    0.5-1.0$, i.e. apastron vs. $[(t-t_0)/1626]$. Data up to MJD~56838
    (the time interval in \citealt{Jaron2014}) are plotted in black;
    data newer than that appear in red. The long-term amplitude
    modulation is clearly visible. The two grey areas indicate the
    $\Theta$ intervals that are removed from the data for the analysis
    of this paper.
  }
  \label{fig:FermiLATaa}
\end{figure}

Which data are causing this decline in the power of $P_2$?
In order to find out we remove a certain interval of the long-term
modulation from the data in the sense that we delete data for which
$\Theta = 0.0-0.1,\,0.1-0.2,\,\cdots\,,0.9-1.0$ and perform
Lomb-Scarge timing analysis on the remaining data. In
Fig.~\ref{fig:searchP2} the resulting powers of $P_1$ and $P_2$ are
plotted for the different $\Theta$ intervals removed from the
data. While the power of $P_1$ is only slightly affected by the choice
of the removed $\Theta$ interval  in a random manner, the power of
$P_2$ is clearly dependent on the removed $\Theta$ interval in a
stronger and systematic way. We find that $P_2$ gains the most
power if we remove the data from the interval $\Theta = 0.5-0.6$, which
is indicated by the grey areas in Fig.~\ref{fig:FermiLATaa}, where we
plot all \fermilat{} data from the orbital phase interval $\Phi =
0.5-1.0$ (i.e. apastron) and averaged over one orbit. The resulting
periodogram, plotted in Fig.~\ref{fig:timing}\,c, shows two peaks at
$P_1$ and $P_2$ with equal power.

Something in the interval around $\Theta \sim 0.5$ affects the
precessing jet and consequently the timing analysis. In the interval
2006--2007, which corresponds to $\Theta_{\rm integer} = 6.38 - 6.60$,
there was strong (15\%-20\% Crab nebula) TeV flaring activity (see
\citealt{Aliu2013} and references therein) as in 2014 (corresponding
to $\Theta_{\rm integer} = 8.36$ ($> 25\%$ Crab nebula)
\citealt{Archambault2016}). TeV flaring activity was found to be
consistent with the long-term period \citep{ahnen16}. The GeV flares
we found perturbing the precessing jet and likely related to the TeV
flares should then follow the same periodicity. This finding suggests
a link to be explored between the high energy phenomena and the radio
transient in \lsi{}. This bursting phase of optically thin emission
present in \lsi{} as in other microquasars \citep{Massi2014,
  Zimmermann2015} is in fact associated with shocks travelling in the
jet some days after the optically thick outburst of the steady jet
\citep[][Fig.~3]{Massi2014b}.

\section{Results}
\label{sect:results}

In this section we compare the observations with model data calculated
by our physical model of a precessing ($P_2$) jet periodically ($P_1$)
refilled in two different parts of the orbit, with relativistic
electrons and emitting synchrotron emission in the radio band (Eq.~1
in \citealt{Massi2014}) and inverse Compton emission in the GeV band
(Eq.~{\ref{eq:flux}). Model data were fitted to the observations
  versus time with the additional constraint that the periodograms of
  model data have to reproduce the periodograms of the
  observations. The fit results for the two ejections are given in
  Table~1. The resulting fit results for the jet are given in Table~2.

Figure~\ref{fig:losses} shows the resulting shape of the two
electron distributions injected around periastron  and
towards apastron. The relativistic electron density\ $K_{\rm
  I}$ is larger than $K_{\rm II}$ as for the accretion peaks of Fig.~6
in \citet{Marti1995}. The resulting mean energy injected per second in
the form of relativistic electrons is $L_{\rm rel} =
8~10^{35}$~ergs/sec. This value is consistent
with the observed gamma-ray luminosity,  $7~10^{35}$~erg/sec,
resulting from the \fermilat{} data.

The resulting jet lengths, $L$, are derived by the analysis of the
energetic losses of the electrons; Figure~3~b reports how the maximum
electron Lorentz factor evolves along the jet axis and where it
attains a value that is too low to be able to produce radio emission.

\begin{table}
  \centering
  \begin{tabular}{lllll}
    \hline
    \hline
    & $\Phi_0$        & $n$          & $ex$  & $A / 10^6$ \\
    \hline
    I  & $0.35 \pm 0.01$         & $0.4 \pm 0.1$         & $4
    \pm 2$ & $25 \pm 1.3 $\\
    II    & $0.58 \pm 0.03$ & $5.5 \pm 2.5$ & $1.95 \pm 1.05 $ &
    $5 \pm 0.6$\\
    \hline
    \hline
  \end{tabular}
  \caption{
    Parameters for the distributions of injected relativistic electrons
    (Fig.~3\,a)
  }
  \label{tab:fit}
\end{table}

\begin{table}
  \centering
  \begin{tabular}{lll}
    \hline
    \hline
    & $\beta$ & $L$ \\
    \hline
    I  & $0.006 \pm 0.004$ & $1.20 \pm 0.01$ \\
    II    & $0.5 \pm 0.2$   & $25 \pm 2$ \\
    \hline
    \hline
  \end{tabular}
  \caption{
    Jet parameters. The position of the base of the jet, $x_0$, and
    the initial magnetic field strength are the same at all orbital
    phases: $x_0 = 3.8 \pm 0.1\,10^{12}$~cm, $B_0 = 5.1 \pm
    0.1$~G. The maximum Lorentz factor was set to $\gamma_2 = 5 \pm 2
    \,10^4$ (Sect.~3).
   }
   \label{tab:results}
\end{table}

We performed timing analysis of \fermilat{} observations and our model
data; Fig.~\ref{fig:timing} shows the results. The observed
$\gamma$-ray flux is modulated by only the orbital period\ $P_1$ at
periastron (Fig.~\ref{fig:timing}\,a). Also the model data are only
modulated by the same period\ $P_1$ (Fig~\ref{fig:timing}\,b). At
apastron, both observational and model data are not only modulated by
$P_1$ but also by $P_2$ (Fig~\ref{fig:timing}\, c and d). Radio
observations and model radio data both show the two features at
$P_1$ and $P_2$ (Fig~\ref{fig:timing}\, e and f).

\begin{figure*}
  \includegraphics{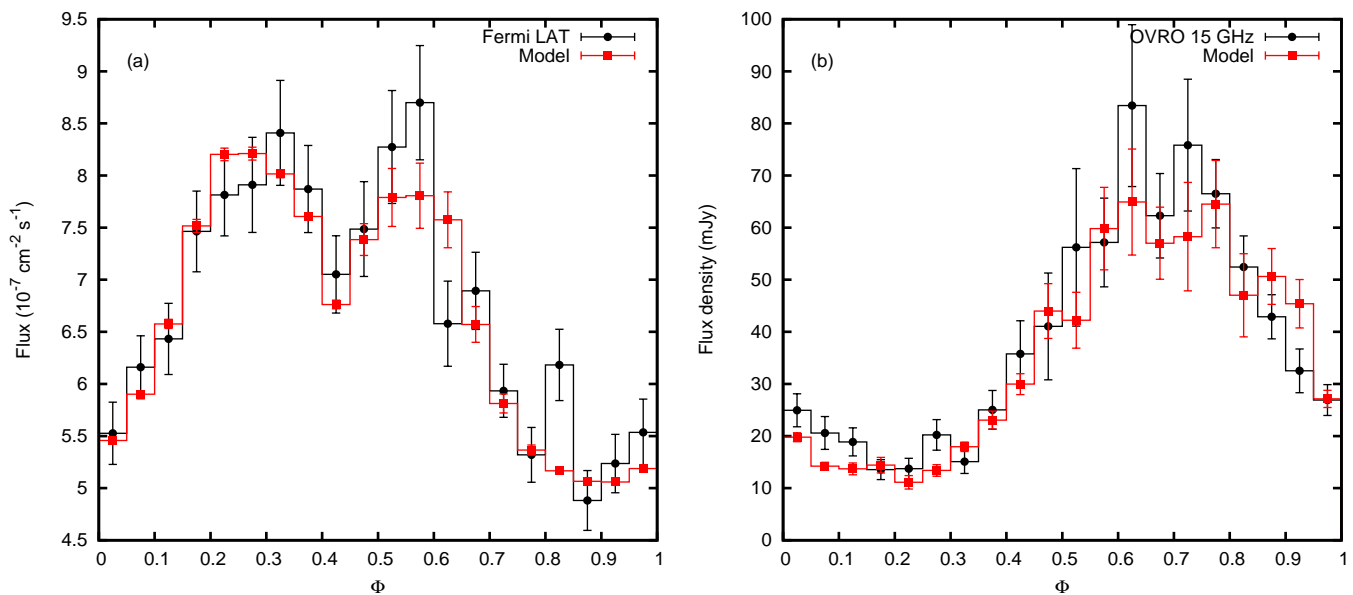}
  \caption{
    Results of the model in comparison with the observations.
    (a) Comparison of the model GeV output (red) with the observations
    by the \fermilat{} (black), the energy range is 0.1--3.0~GeV in
    both cases and the time interval is MJD~54682--57450. The observed
    \fermilat{} light curve, here folded on the orbital period $P_1 =
    26.4960$~days, shows two peaks. Assuming two electron injections
    along the orbit, the shape of the orbital  modulation of this
    observed GeV light curve is reproduced.
    (b) Comparison of the model radio output (red) with the
    observations by OVRO (black), both at 15~GHz spanning the time
    interval MJD~54909--56794. The observed light curve, folded on the
    orbital period $P_1$, shows only one peak along the orbit.
     The model
    reproduces this characteristic apastron peak even though two
    electron injections are taken into account. 
  }
  \label{fig:folded}
\end{figure*}

The folded light curves for radio (OVRO) and $\gamma$-ray data
(\fermilat{}) are shown in Fig.~\ref{fig:folded}. Even if $K_{\rm I}$
is larger than $K_{\rm II}$ (Fig.~\ref{fig:losses}), in the right
panel of Fig.~\ref{fig:folded} one sees only the large radio outburst
at apastron for both observations and the model data.

Concerning \fermilat{} data, the two-peaked structure is present in
both observations and model. The two-peaked structure is above an
offset of $\unit[5]{10^{-7} cm^{-2} s^{-1}}$. The origin of the offset
could be SSC of optical photons produced not in the jet but in the hot
flow that fills the inner part of the accretion flow. Indeed, optical
observations by \citet{Casares2005} indicated an  extra contribution
to the stellar emission that the authors  attributed possibly to the
disk around the Be star. Today it is suggested that in X-ray binaries
optical synchrotron radiation might come from  the hot accretion flow
\citep{Poutanen2014}. The  SSC emission of this optical radiation,
invoked to explain the MeV tail in the hard state of Cygnus X-1
\citep{Poutanen2014}, seems to extend to GeV here, as the \lsi{}
spectrum from 0.01--1000~MeV in Fig.~5 by \citet{Tavani1996}
suggests. Optical observations by \citet{Mendelson1989} measured two
variations at periastron and towards apastron. Our left panel of
Fig.~7, which shows two peaks above the large offset, fits and
completes a scenario in which the offset could be associated with the
SSC of optical photons of the hot accretion flow and the two peaks to
SSC and EIC of jet and stellar photons scattered by electrons of the
jet. Concerning this jet component, the main contribution comes from
the SSC reprocessing for the parameter range resulting from the above
fit. The relative importance of SSC emission with respect to EIC
emission, even if always in favour of the former, changes not only
with the orbital phase $\Phi$, but also shows a long-term modulation
with $\Theta$.  

\section{Conclusions}
\label{sect:conclusions}

The open issue this paper aimed to address is why the long-term
modulation does not affect the gamma emission of \lsi{} at periastron,
or equivalently why $P_2$ is missing in the periodogram of the
gamma-ray emission at periastron. In addition, we wanted to understand
whether there is any relationship between this lack of $P_2$ in the
gamma-ray emission and the absence of a radio outburst at periastron.

We modelled the observed flux in terms of synchrotron and IC emission
of a precessing jet (with period $P_2$) periodically ($P_1$) refilled,
twice along the orbit, with relativistic electrons.

The model solution implies that the ejection at periastron, where the
compact object is embedded in the densest part of the Be wind, is
indeed the larger of the two ejections. The larger inertia, because
of the higher accretion rate and the dense environment where the jet
propagates, may explain why the jet at periastron is
slow; the model solution gives at periastron $\beta = 0.006 \pm
0.004$.

This slow dense jet, which at periastron is embedded in the stellar
photon field, suffers from strong energetic losses via IC,
i.e.\ electrons upscatter optical/UV photons to gamma-ray energy and
by doing that quickly lose their energy. This explains why at
periastron there is strong gamma-ray emission and only negligible
radio emission. In fact, we find that the electrons in the jet at
periastron reduce their Lorentz factor under the level at which they
can emit radio (15~GHz) synchrotron radiation already at the very
short distance of $L = 1.20 \pm 0.01$. That corresponds to $x = Lx_0 =
3.0\cdot10^6r_{\rm s}$ (for $M_{\rm BH} = 3M_\odot$), that is barely
after the acceleration/collimation region of a jet that is at $\sim
10^{5.5} r_{\rm s}$ \citep{Marscher2006}. For the second minor
ejection towards apastron the length of the jet results to be $L = 25
\pm 2$, i.e. $x = 6.3 \cdot10^7r_{\rm s}$ and $\beta = 0.5 \pm 0.2$.

The observed flux is  $S_0(P_1, P_2) = S_{\rm i}(P_1){\rm
  DB}(P_2)$. The intrinsic flux\ $S_{\rm i} = f(P_1)$ is due to IC and
synchrotron processes. The  DB is a function of $\beta
\cos(\eta(f(P_2)))$. In fact, the precession of the jet periodically
changing the angle\ $\eta$ between the jet axis and line of sight
changes the DB. The low velocity at periastron gives rise to a small
Doppler factor and hence to a negligible dependency on $P_2$. As a
result, the $\gamma$-ray flux at periastron is neither modulated by
the long-term periodicity \citep{Ackermann2013} owing to the beating
between $P_1$ and $P_2$ \citep{Jaron2014} nor does $P_2$ appear in the
timing analysis. In fact, each period, $P_1$ and $P_2$, only appears
in the timing analysis when the related term ($S_{\rm i} (P_1)$ or
${\rm DB}(P_2)$) is significant. At apastron, the larger velocity
creates larger DB variations, $P_2$ appears in the timing analysis,
and radio and gamma-ray emission show the long-term periodicity.

\begin{acknowledgements}
  We thank Eduardo~Ros and Karl~Menten for reading the manuscript. We
  thank Bindu~Rani
  and Robin~Corbet for useful discussions about likelihood analysis of
  \textit{Fermi}-LAT data. We thank Alan~Marscher for useful
  discussions about jet acceleration. Helge Rottmann provided us with
  computing power. This work has made use of public
  \textit{Fermi} data obtained from the High Energy Astrophysics
  Science Archive Research Center (HEASARC), provided by NASA Goddard
  Space Flight Center. The OVRO 40~m Telescope Monitoring Program is
  supported by NASA under awards NNX08AW31G and NNX11A043G, and by the
  NSF under awards AST-0808050 and AST-1109911.
\end{acknowledgements}

\bibliographystyle{aa}

\end{document}